\documentstyle[preprint,aps,epsf]{revtex} 
\tighten 
 
\begin{document} 
 
\draft 
 
\title{Quantized bulk fermions in the Randall-Sundrum brane model} 
 
\author{Antonino Flachi\thanks{e-mail address:{\tt
antonino.flachi@ncl.ac.uk}}, Ian G. Moss \thanks{e-mail address:{\tt  
ian.moss@ncl.ac.uk}} and David J. Toms\thanks{e-mail address:{\tt
d.j.toms@ncl.ac.uk}}}
\address{Department of Physics, University of Newcastle Upon Tyne,
NE1 7RU United Kingdom}

\date{\today} 
 
\maketitle 
 
\begin{abstract} 
The lowest order quantum corrections to the effective action arising
from  
quantized massive fermion fields in the Randall-Sundrum background  
spacetime are computed. 
The boundary conditions and their relation with gauge invariance 
are examined in detail.  
The possibility of Wilson loop symmetry breaking in brane models  
is also analyzed. 
The self-consistency requirements, previously considered in the case of
a quantized bulk scalar field,  
are extended to include the contribution from massive fermions. It is
shown that in this case it is possible to stabilize the radius of the
extra dimensions but it is not possible to simultaneously solve the hierarchy
problem, unless the brane tensions are dramatically fine tuned, supporting
previous claims. 
\end{abstract}  
\vskip0.5cm 
PACS number(s):11.10.Kk, 04.50.+h, 04.62.+v, 11.25.Mj 
\eject 
 
\def\SS{\Sigma_{IJ}^S} 
\def\SF{\Sigma_{IJ}^F} 
\def\DS{\Delta_{IJ}^S} 
\def\DF{\Delta_{IJ}^F} 
\def\O{\Omega} 
\def\L{\Lambda} 
 
 
\section{Introduction} 
 
The idea of extra dimensions was originally introduced in order to  
provide a unified description of the electromagnetic and gravitational  
interactions \cite{Kaluza,Klein} and further generalised to more than
one extra dimension in  
order to allow the incorporation of non-abelian gauge fields
\cite{DeWitt}.  
More recent interest was generated in connection with supergravity and  
string theory \cite{Witten,Nahm,DuffNilssonPope}. 
 
In the past few years this idea is having a novel rejuvenation,  
particularly in relation with the resolution of the hierarchy problem. In
addition, extra dimensions 
have also provided an interesting link between string theory and
particle  
physics, motivating the construction of low energy effective theories
with possible experimental  
signatures, and suggesting possible resolutions of many long standing  
problems of particle physics and cosmology. 
 
This new perspective on higher dimensional theories was first pointed
out in \cite{Arkani}, where,  
in contrast to the standard belief that extra dimensions must be  
associated with extremely small length scales, it was noted that the
extra  
dimensions could be as large as a millimeter, bringing the fundamental
Planck scale  
closer to the electroweak scale and thus providing an explanation for
the relative weakness of  
gravity with respect to the other forces.  
Unfortunately, this scenario with large extra dimensions suffers from
an important drawback. 
It trades, in fact, a large ratio between the Planck scale and the
electroweak scale  
for a large ratio between the compactification scale and the electroweak  
scale, not providing a satisfactory explanation to the hierarchy problem. 
 
A brane model, with the interesting feature of having all the  
parameters of the theory of the same magnitude while still  
generating a very large hierarchy, was devised by Randall and Sundrum  
\cite{RS}.  
Their model is based on a 5-dimensional spacetime  
with the extra spatial dimension having an orbifold compactification. Two
 
3-branes with opposite tensions sit at the orbifold fixed points. The
line  
element is 
\begin{equation} 
ds^2=e^{-2kr|\phi|}\eta_{\mu\nu}dx^\mu dx^\nu-r^2d\phi^2\label{1.1} 
\end{equation} 
with $x^\mu$ the usual 4-dimensional coordinates, $|\phi|\le\pi$ with
the  
points $(x^\mu,\phi)$ and $(x^\mu,-\phi)$ identified.  
of  
The 3-branes sit at  
$\phi=0$ and $\phi=\pi$. $k$ is a constant of the order of the Planck  
scale (the natural scale for the theory), and $r$ is an arbitrary
constant  
associated with the size of the extra dimension. 
The interesting feature of the Randall-Sundrum model is that it can  
generate a TeV mass scale from the Planck scale in the higher
dimensional  
theory.  
A field with a mass $m_0$ on the $\phi=\pi$ brane will have a  
physical mass of $m\simeq e^{-\pi kr}m_0$. By taking $kr\simeq12$, and  
$m_0\simeq10^{19}$~GeV, we end up with $m\simeq1$~TeV.  
 
Another interesting aspect of brane models is related to their field
content. 
In the original version of the Randall-Sundrum model all of the standard  
model particles were confined on the brane, with only gravity moving  
throughout the bulk spacetime. Alternatives to confining particles on
the  
brane have been investigated and a different number of reasons seem to  
suggest the necessity of new bulk physics  
\cite{GW1,DHR,Pomarol,Grossman,Kitano,Chang,Gherghetta,Changetal,DHR2,Huber,FTplb}.
 
Particularly relevant is the role of higher dimensional bulk fermions,
primarily  
in relation with string theory, as they arise as superpartners of
gravitational  
moduli and are inevitable in any string theory realisation of the brane
world idea. 
In the context of  particle physics phenomenology and string inspired
model  
building some study has been devoted to include bulk fermions, but apart
from a few exceptions,  
attention has been mostly concentrated on the massless case. However
bulk fermion masses constitute  
an important feature which has to be taken into account for a different
number of reasons. 
 
In order to study possible phenomenological signatures, massless
\cite{Changetal} and massive \cite{DHR2} 
bulk fermions have been considered. Interestingly, in \cite{DHR2}  
the resulting phenomenology is shown to be highly dependent on the value
of the five dimensional fermion mass. In \cite{Grossman} a new way for  
obtaining small Dirac neutrino masses, without invoking a see-saw  
mechanism, was outlined. Within an effective field theory approach,  
massless chiral fermions and loop corrections to the effective action
have  
been investigated \cite{georgi1,georgi2}. In \cite{lukas} a comprehensive
study of five dimensional brane models for neutrino physics has been  
presented.  
 
Motivation for introducing massive bulk fermions also come from the need
to localise fermion zero modes in the extra dimensions. 
In \cite{daemi} a modification of the Dirac equation via a pseudo-scalar
Yukawa coupling term of the form 
$m \chi(y)\bar{\Psi} \Psi$, with $\chi \propto \epsilon(y)$ and
$\epsilon(y)$ the sign function,  
has been considered and it was shown that in this way it is possible to
ensure both localisation and chirality. Non-chiral theories of
fermions have been discussed in \cite{FMTPLB}, where we stressed the
fact that the fermion representations of the full Lorentz group in
five dimensions have eight, rather than four, components.
 
The radius $r$ of the extra dimensions is assumed to be the vacuum  
expectation value of a scalar field, called the radion. 
In the scenario proposed by Randall and Sundrum, the radion has zero  
potential and consequently $r$ is not determined by the dynamics of the  
model.  
Therefore for this scenario to be physically acceptable, it is necessary  
to find a mechanism for generating such a potential which would
stabilize  
the size of the extra dimensions. 
 
Goldberger and Wise have suggested a solution to this problem
\cite{GW1}.  
They proposed the introduction of a bulk scalar field with appropriate  
interaction terms on the branes as a means to induce a stabilizing
potential 
\footnote{Note that the Goldberger and Wise  
model has to include the backreaction on the metric and the fine tuning
of the  
cosmological constant, in order to satisfy the consistency conditions  
derived in \cite{gibbons}}. Although this model provides a solution to
the problem,  
it can be viewed as being very artificial and hence it is important to  
seek more natural alternatives.  
 
The older Kaluza-Klein theories,  
based on factorisable geometries,  
homogeneous space)  
were affected by a similar difficulty. In that context, it was realised
by Candelas and Weinberg that quantum  
effects from matter fields or gravity could be used to fix the size of
the extra  
dimensions, stimulating the study of quantum effects in such scenarios  
\cite{CW,DJTgravity,DJTChalkRiver,ACFbook}. 
Analogously to that example it seems reasonable to investigate if the
radius of  
the extra dimensions can be determined by quantum effects.  
Motivated by this analogy, the role of quantum effects has received some  
recent attention \cite{Garriga,DJTPLB,GR,flachi,FMTPLB,ponton,horava}.  
 
In \cite{Garriga} massless and conformally coupled fields obeying
untwisted boundary conditions  
have been analysed. Massive scalar fields minimally coupled to the
scalar  
curvature with untwisted boundary conditions have been considered in  
\cite{DJTPLB,GR}. In \cite{flachi} massive scalar fields obeying twisted
and untwisted boundary conditions with a  
non-minimal coupling have been investigated and a self-consistency
relation has been obtained.  
In \cite{flachi} the importance (in principle) for the inclusion of the
induced gravity term has been  
pointed out and it has been computed in both the twisted and untwisted
case. Massless fermion fields have  
been investigated in both \cite{Garriga} (for untwisted field  
configurations) and in \cite{FMTPLB} (for twisted as well as untwisted  
field configurations). Also in \cite{FMTPLB} it has been pointed out
that the boundary condition  
structure it can be enlarged when considering a gauge symmetry. 
The main conclusion of \cite{Garriga,DJTPLB,GR,flachi,FMTPLB} is that
a severe fine tuning of the brane  
tensions is essential for the radius to be stabilized via quantum effects
and the  
hierarchy problem to be solved simultaneously.  
The one-loop Casimir energy in five dimensional $S^1/{Z}_2$
and six dimensional  
$T^2/{Z}_k$ orbifolds has been considered in \cite{ponton}. 
Some related work in M theory has been done in \cite{horava}, where the
Casimir energy is evaluated for a  
non-supersymmetric $E_8 \times \bar{E}_8$ compactification of M theory
on $S^1/{Z}_2$.

In the present paper we try to amplify the previous considerations and
compute  
the radiative corrections to the effective action on a five  
dimensional classical background spacetime with an orbifold
compactification including the contribution  
coming from massive Fermi fields. 
The next section is devoted to introducing the general framework and
compute the one-loop vacuum energy  
for a single 4-component fermion. The relation between the boundary
condition structure and gauge invariance  
is clarified in section 3 and the vacuum energy is computed for a fermion
and scalar multiplet under general  
boundary conditions. The massless, conformally coupled case is discussed
in section 4, as it provides a useful check  
on the method used. 
The possibility of Wilson loop symmetry breaking is deferred to section
5.  
In section 6 we discuss the self-consistency requirements in the model
when quantum effects  
are included. Our conclusions are drawn in the last part of the paper. 
 
 
\section{Effective Action} 
 
In this section we will evaluate the quantum corrections to a classical  
theory specified by a single massive bulk fermion on the Randall-Sundrum  
background spacetime. 
We follow the general method outlined in \cite{DJTPLB}, which we will
briefly  
review. 
It consists of first expanding the higher dimensional fields in terms
of a complete  
set  of modes and then integrating out the dependence on the extra
dimension  
leaving an equivalent four dimensional theory with an infinite number
of fields  
with masses quantized in some way: 
\begin{equation} 
	S = \sum_n S_n~. 
\end{equation} 
$S_n$ represents the dimensionally reduced theory for the $n^{th}-$mode,
with differential operator  
given by ${\cal D}_n$. Having done this, the effective action is simply
given by the sum over the modes: 
\begin{equation} 
	\Gamma^{(1)}_n = {1 \over 2} \sum_n  \ln det \left( \ell^2 {\cal
	D}_{n}  
	\right)~. 
\end{equation} 
Since the one-loop effective action is expressed as the logarithm of  
the determinant of ${\cal D}_n$, it turns out to be advantageous to
adopt a  
heat kernel method, namely writing the effective action in terms of a  
non-local kernel function $K_n (s,x,x')$: 
\begin{equation} 
	\Gamma^{(1)} = -{1 \over 2} \sum_n \int d^Dx |g|^{1/2} \int {ds
	\over s} 
	Tr K_n(s,x,x)~, 
\end{equation} 
where 
\begin{equation} 
	- {\partial \over \partial s} K_n(s,x,x') = {\cal D}_n
	K_n(s,x,x')~, 
\end{equation} 
\begin{equation} 
	K_n(0,x,x') = \delta (x,x')~. 
\end{equation} 
It is now possible to use an asymptotic expansion for the heat kernel,
in order to obtain an expansion in powers of the curvature: 
\begin{equation} 
	K_n(s,x,x) \propto s^{-D/2} \sum_{k=0}^\infty (is)^k a_k(x)~, 
\end{equation} 
where the heat kernel coefficients $a_k(x)$ depend on geometrical
invariants only. 
 
The coefficients $a_k(x)$ are known for a wide class of differential
operators defined on manifolds with and without  
boundaries with different types of boundary conditions (see
\cite{DeWitt,Gilkey1,Gilkey2,kirsten} and references therein), and  
recently the spectral geometry of operator of Laplace type on manifolds
with singular surfaces has been considered also in  
connection with brane models \cite{Gilkey3,mossplb,bordag}.\\ 
The leading term, given by $a_0(x)$, represents the Casimir energy
contribution to the  
cosmological constant. The next term is  
proportional to the four dimensional curvature and gives a gravity term  
induced by quantum effects. This term has received little attention in  
brane models, but it played a major role in the study of the  
self-consistency of the older Kaluza-Klein theories
\cite{kunst}. Additionally,  
the induced gravity term is essential if we wish to identify the
physical  
value of the Newtonian gravitational constant in terms of the bare one. 
The next term in the  
expansion, proportional to $a_3(x)$, contains higher curvature terms and  
becomes important when considering higher derivative gravity models on
brane backgrounds  
\cite{kim1,kim2,corradini,nojiri,navromatos,neupane,low}. 
 
Strictly speaking, in the Randall-Sundrum scenario, where the 3-branes
are  
flat, these terms are absent, however they make their appearance when  
curved branes are considered. 
In our analysis, we are simply after the vacuum energy and therefore
these  
next-to-leading terms won't be reported, although they are obtainable at  
ease with simple modifications of our calculation. 
 
The result for the quantum corrections to the effective action is found
to be  
divergent and needs to be regulated in some way. Here, following the  
procedure of \cite{flachi}, we choose to use dimensional regularisation. 
 
 
\subsection{Kaluza-Klein reduction} 
 
The Kaluza-Klein reduction has been performed in a number of references
(see for example  
\cite{Grossman}) and the details won't be repeated here. Only in order
to fix the notation and discuss  
few points of importance for the present work, we outline the essential  
steps. 
Initially, we consider a single chiral fermion on the spacetime described by
(\ref{1.1}) and whose action is given by 
\begin{equation} 
	S =  \int d^4x \int dy |g|^{1/2} \left( i \bar{\Psi}
	\underline{\gamma}^M  
	D_M \Psi - m\cdot \epsilon(y) \bar{\Psi} \Psi \right)~. 
	\label{5daction} 
\end{equation} 
For notational convenience, the coordinate $y$ describing the extra
dimension is reparametrised by 
\begin{equation} 
	y=r \phi~, 
\end{equation} 
The Kaluza-Klein reduction can be performed by decomposing the field
$\Psi$ in its left and right components, 
\begin{equation} 
	\Psi(x_\mu, y) = \sum_n \left( \Psi_R^{(n)}(x_\mu) g_R^{(n)}(y)
	+   
	\Psi_L^{(n)}(x_\mu) g_L^{(n)}(y) \right) 
\end{equation} 
and then integrating over the extra dimension. This leaves us with 
\begin{equation} 
	S = \sum_n \int d^4x \left(  
	i \bar{\psi}_n \mbox{${\not{\hbox{\kern-2.0pt$\partial$}}}$}\psi_n
	 	- m_n \bar{\psi}_n \psi_n \right)~, 
	\label{4daction} 
\end{equation} 
where $\psi_n(x) = \Psi_L^{(n)}(x) + \Psi_R^{(n)}(x)$. The dependence
of $\Psi$   
on the extra dimension can be expressed as a combination of Bessel  
functions: 
\begin{equation} 
G_L^{(n)}(z_n) = z_n^{1/2} \left( a^{(n)}_L J_{1/2 + \nu} (z_n) +
b^{(n)}_L  
J_{-1/2 - \nu} (z_n)\right)~, 
	\label{gggl} 
\end{equation} 
\begin{equation} 
G_R^{(n)}(z_n) = z_n^{1/2} \left( a^{(n)}_R J_{1/2 - \nu} (z_n) +
b^{(n)}_R  
J_{-1/2 + \nu} (z_n)\right)~, 
	\label{gggr} 
\end{equation} 
where $G_{(R,L)}^{(n)} = e^{-2k|y|} g_{(R,L)}^{(n)}$, $z_n = m_ne ^{k|y|}/k$
and   $\nu = m/k$. 
 
The coefficients $a^{(n)}_L$, $b^{(n)}_L$, $a^{(n)}_R$, $b^{(n)}_R$
can be found  
by imposing some boundary conditions. The possible boundary conditions
are related to  
the parity of the spinor field $\Psi$ under a chiral transformation. A
possibility,  
which we will call $I$, is that the field $\Psi$ is even: 
\begin{equation} 
	\gamma_5 \Psi(x_\mu, -y) = +\Psi(x_\mu, y)~, 
\end{equation} 
implying that the mass eigenvalues are quantized according to the
following transcendental equation: 
\begin{equation} 
	J_{1/2 + \nu} ({m_n \over ka}) J_{-1/2 - \nu} ({m_n \over k}) -  
	J_{1/2 + \nu} ({m_n \over k}) J_{-1/2 - \nu} ({m_n \over ka})
	= 0~, 
\end{equation} 
where $a=e^{-kr\pi}$. 
The other possibility, which we will call $II$, is given by: 
\begin{equation} 
	\gamma_5 \Psi(x_\mu, -y) = -\Psi(x_\mu, y)~,  
\end{equation} 
leading to  
\begin{equation} 
	J_{-1/2 + \nu} ({m_n \over ka}) J_{1/2 - \nu} ({m_n \over k}) -  
	J_{-1/2 + \nu} ({m_n \over k}) J_{1/2 - \nu} ({m_n \over ka})
	= 0~. 
\end{equation} 
It is worth commenting a bit further on the two types of parity
conditions, and the mass term chosen in  
(\ref{5daction}). Generally we would expect that because of the
${Z}_2$ identification of the extra dimension, we could have  
\begin{equation} 
	\Psi(x,-y) = B \Psi(x,y) 
	\label{bubu} 
\end{equation} 
for some matrix $B$. The requirements that the action, or Hamiltonian,
remain invariant under the ${Z}_2$  
identification place certain constraints on $B$, which are easily shown
to be 
\begin{equation} 
	B^\dagger B = I 
\end{equation} 
\begin{equation} 
	[\gamma^0 \gamma^i , B] = 0,~\mbox{$i=1,~2,~3$} 
\end{equation} 
\begin{equation} 
	\{\gamma^0 \gamma_5 , B\} = 0 
\end{equation} 
where $\{,\}$ is the anticommutator. The only way to solve these relations
occurs if  
\begin{equation} 
B=e^{i \delta} \gamma_5 
\end{equation} 
for some arbitrary phase factor $\delta$. Finally we use the fact that
$B$ as defined in  
(\ref{bubu}) must provide a representation of the group ${Z}_2$,
which requires  
\begin{equation} 
B^2 =I~. 
\end{equation} 
(This is just a fancy way of saying that two reflections gives us back
the  
identity.) This fixes $\delta$ to be $0$ or $\pi$, and hence 
\begin{equation} 
B=\pm \gamma_5 
\end{equation} 
are the only two possibilities.  
 
Regarding the mass term, the identification of $y$ with $-y$ on the
fields with $B=\pm \gamma_5$ does not leave  
the mass term invariant if $m$ is a constant. Choosing $m \propto
\epsilon(y)$ is the simplest possibility for an invariant mass term.
(A constant mass term can be used with eight component spinor
representations \cite{FMTPLB}). 
 
\subsection{Evaluation of the vacuum energy} 
 
We want to compute the vacuum energy for the theory described by the
action (\ref{4daction}).  
The fact that type $I$ and type $II$ boundary conditions differ only
for the order of the Bessel functions  
simplifies the subsequent analysis and allows to deal with both cases
at once. In what follows we define 
\begin{equation} 
	\mu = 1/2 + \nu, \mbox{for type $I$ boundary conditions} 
\end{equation} 
and 
\begin{equation} 
	\mu = 1/2 - \nu, \mbox{for type $II$ boundary conditions}~. 
\end{equation} 
Following \cite{flachi}, we use the form for the heat kernel described
in \cite{parker}. 
After some well known manipulation the one-loop effective action can be
expressed as 
\begin{equation} 
	\Gamma^{(1)} = -{i\over 2} \sum \ln det \left( \ell^2( \Box
	+{1\over 4}R  
	+m_n^2)\right)~. 
\end{equation} 
Now using the heat kernel expansion for the previous operator we have: 
\begin{equation} 
	\Gamma^{(1)} = \int d^4x |g|^{1/2} {\cal L}_\Lambda^F~, 
\end{equation} 
with 
\begin{equation} 
    {\cal L}_\Lambda^F = -{1\over 2}(4\pi)^{-2} \lim_{D\rightarrow 4}
    \Gamma (-D/2)\sum_n  
    m_n^D~. 
\end{equation} 
The masses $m_n$ are quantized according to the following eigenvalue
equation: 
\begin{equation} 
    P_\mu (x_n) =	J_{\mu} (x_n) J_{-\mu} (a x_n) -  
	J_{\mu} (a x_n) J_{-\mu} (x_n) = 0~, 
	\label{quantmass} 
\end{equation} 
where, for convenience, we have defined 
\begin{equation} 
	m_n = k a x_n~. 
\end{equation} 
Introducing $\epsilon = D-4$, we find 
\begin{equation} 
    {\cal L}_\Lambda^F = -{1\over 32\pi^{2}} \lim_{\epsilon \rightarrow 0}
    (ka)^{4+\epsilon}  
    v(\epsilon)~, 
\end{equation} 
where 
\begin{equation} 
    v(\epsilon) = \Gamma (-2-\epsilon/2) \sum_n x_n^{4+\epsilon}~. 
\end{equation} 
The method we use to compute the previous sum is a simple modification
of  
the technique developed in \cite{kirst1,kirst2}, which allows to evaluate
the $\zeta -$function  
using only the basic properties of the eigenvalue equation. 
 
A simple application of the residue theorem permits to convert the
previous sum into a contour integral: 
\begin{equation} 
v(\epsilon)=\frac{\Gamma(-2-{\epsilon \over 2})}{2\pi i}\oint_{\mathcal  
C}dz\,z^{4+\epsilon}\,\frac{d}{dz}\ln \,P_\nu(z)\;, 
\end{equation} 
where ${\mathcal C}$ is any contour which encloses the positive zeros
of $P_\nu(z)$.  
After some manipulations, with an appropriate choice for the contour
${\mathcal C}$,  
$v(\epsilon)$ can be recast in the following form:  
\begin{equation} 
    v(\epsilon) = {1 \over \Gamma (3+\epsilon/2)} \int_0^\infty dy
    y^{4+\epsilon}  
    {d\over dy} \ln Q_\mu (y)~, 
\label{int} 
\end{equation} 
with  
\begin{equation} 
    Q_\mu (y) = I_\mu (y) K_\mu (ay) - I_\mu (ay) K_\mu (y)~. 
\end{equation} 
Expression (\ref{int}) can be rearranged in order to isolate the
divergent  
contributions and exploiting the dependence on $a$. The analytical
continuation of  
(\ref{int}) to $\Re(\epsilon)>2$ can be carried out noting that the
impediment to the convergence of (\ref{int}) comes  
from the behaviour of $Q_\mu(z)$ at large $z$. 
Analogously to the scalar field case, we will define 
\begin{equation} 
I_\nu(z) = \frac{e^z}{\sqrt{2\pi z}}\,\Sigma^{(I)}(z)\;, 
\end{equation} 
\begin{equation} 
K_\nu(z) = \sqrt{\frac{\pi}{2 z}}\,e^{-z}\,\Sigma^{(K)}(z)\;. 
\end{equation} 
For large $z$ the asymptotic expansions for the Bessel functions show
that 
\begin{equation} 
\Sigma^{(I)}(z)\simeq\sum_{k=0}^{\infty}\alpha_k\,z^{-k}\;, 
\end{equation} 
where 
\begin{equation} 
\alpha_k=\frac{(-1)^k\,\Gamma(\nu+k+\frac{1}{2})}
{2^k\,k!\,\Gamma(\nu-k+\frac{1}{2})}\;,
\end{equation} 
and 
\begin{equation} 
\Sigma^{(K)}(z)\simeq\Sigma^{(I)}(-z)\;. 
\end{equation} 
With these positions, after some manipulations of (\ref{int}), we end
with 
\begin{equation} 
		v(\epsilon) = J(a) + A + a^{-4 -\epsilon} B -{4d_4 \over
		\epsilon  
		\Gamma(3+\epsilon/2)}(1+a^{-4-\epsilon})~, 
\end{equation} 
where 
\begin{equation} 
	J(a) = {1 \over \Gamma (3+\epsilon/2)} \int_0^\infty dy
	y^{4+\epsilon} 
	{d\over dy} \ln~\left( 1- {K_\mu (y) I_\mu (ay) \over K_\mu (ay)
	I_\mu  
	(y)}\right)~, 
	\label{53} 
\end{equation} 
\begin{displaymath} 
	A =  {1\over \Gamma (3+\epsilon/2)} \left\{ 
	\int_1^\infty dy y^{4+\epsilon} 
	{d\over dy}\left( \ln \Sigma^{(I)}(y) - \sum _{n=1}^N d_n
	y^{-n}\right) \right. 
\end{displaymath} 
\begin{displaymath}	 
	+ \int_1^\infty dy y^{4+\epsilon} 
	{d\over dy}\left( \sum _{n=1, n \neq 4}^N d_n y^{-n}\right) + 
\end{displaymath} 
\begin{equation}	 
	\left. + \int_0^1 dy y^{4+\epsilon} 
	{d\over dy} \ln I_\mu (y) 
	+ \int_1^\infty dy y^{4+\epsilon} 
	{d\over dy} \ln {e^{z} \over \sqrt{2 \pi z}}\right\}~, 
\end{equation} 
\begin{displaymath} 
	B = {1\over \Gamma (3+\epsilon/2)} \left\{ 
	 \int_1^\infty dy y^{4+\epsilon} 
	{d\over dy}\left( \ln \Sigma^{(K)}(y) - \sum _{n=1}^N (-1)^{-n}
	d_n y^{-n}\right)+\right. 
\end{displaymath} 
\begin{displaymath}	 
	+ \int_1^\infty dy y^{4+\epsilon} 
	{d\over dy}\left( \sum _{n=1, n \neq 4}^N (-1)^{-n} d_n
	y^{-n}\right)+ 
\end{displaymath} 
\begin{equation}	 
	\left.+ \int_0^1 dy y^{4+\epsilon} 
	{d\over dy} \ln K_\mu (y) 
	+ \int_1^\infty dy y^{4+\epsilon} 
	{d\over dy} \ln \left( e^{-z} \sqrt{{\pi \over	2 z}} \right)
	\right\}~. 
\end{equation} 
The coefficients $d_n$, which determine the pole structure of the vacuum
energy are given by the large-$z$ behaviour of  
$\Sigma^{(I)}(z)$: 
\begin{equation} 
    \ln\,\Sigma^{(I)}(z)\simeq\sum_{k=1}^{\infty}d_k\,z^{-k}~, 
\end{equation} 
and are immediately evaluated using a Taylor expansion.  
 
\begin{figure}
\begin{center}
\leavevmode
\epsfxsize=20pc
\epsffile{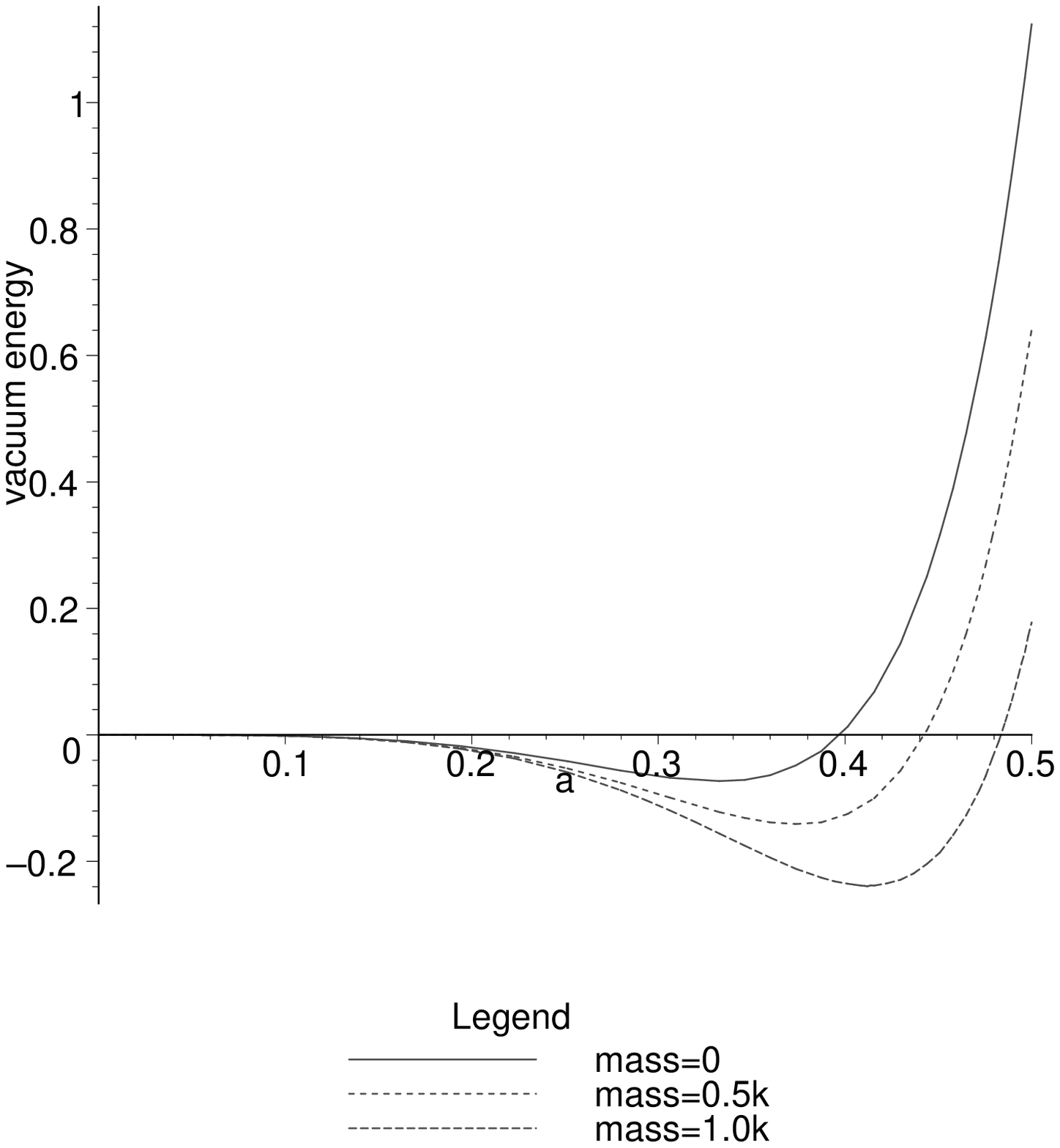}
\end{center}
\caption{The fermion vacuum energy plotted against $a=\exp(-\pi kr)$ for
various fermion masses and type $I$ boundary conditions. The vacuum energy is
in units of $10^3k^4$ and shifted to vanish at $a=0$ to aid comparison. The
brane tension $V_v$ is kept constant.} \protect\label{fig1} 
\end{figure}

The unrenormalized one-loop vacuum energy can, finally, be written as 
\begin{displaymath} 
	{\cal L}_\Lambda^F = -{(ka)^4 \over 64\pi^2} \int_0^\infty dy
	y^{4} 
	{d\over dy} \ln~\left\{ 1- {K_\mu (y) I_\mu (ay) \over K_\mu
	(ay)  
	I_\mu (y)}\right\} -  
\end{displaymath} 
\begin{equation} 
	-{(ka)^{4+\epsilon} \over 32\pi^2} 
	\left(A+a^{-4-\epsilon}B-{4d_4 \over \epsilon  
	\Gamma(3+\epsilon/2)}(1+a^{-4-\epsilon}) 
	\right)~. 
	\label{v0}	 
\end{equation} 
The previous expression is found to be divergent and needs to be
renormalized. 
The renormalization can be performed by using the brane tensions. 
In a previous work \cite{flachi} we have found that when pushing the
heat kernel expansion  
to higher orders there is the need to augment the brane tension with
terms  
proportional to powers of the curvature in order to remove the pole
terms.  
The same happens in the fermion case. 
However, in this calculation we truncated the expansion to first order
and we don't  
need to consider this possibility.  
Proceeding as in the scalar case, one can write the brane sector of the
action as 
\begin{equation} 
	S_{brane} = -\int d^4x|g|^{1/2} 
	\left\{ 
	V_h+a^4V_v 
	\right\}~, 
	\label{branepart} 
\end{equation} 
and express the bare quantities in terms of the renormalized ones: 
\begin{equation} 
	V_{v,h} = V_{v,h}^R + \delta V_{v,h}~, 
	\label{ren1} 
\end{equation} 
where 
\begin{equation} 
	\delta V_{v,h} = {\delta V_{v,h}^{-1} \over \epsilon} +\delta
	V_{v,h}^0~,   
	\label{ren2} 
\end{equation} 
with $\delta V_{v,h}^{-1}$ and $\delta V_{v,h}^{0}$ independent of  
$\epsilon$. 
By using (\ref{v0}), (\ref{ren1}), (\ref{ren2}) the renormalization is  
straightforward leading to the following counterterms: 
\begin{equation} 
	\delta V_h^{-1} = \delta V_v^{-1} = {k^4 \over 16 \pi^2}d_4~,  
	\label{c1} 
\end{equation} 
\begin{equation} 
	\delta V_h^{0} = - {k^4 \over 32 \pi^2} B~, 
	\label{c2} 
\end{equation} 
\begin{equation} 
	\delta V_v^{0} = - {k^4 \over 32 \pi^2} A~. 
	\label{c3} 
\end{equation} 
In performing finite renormalizations, the $a$ dependence of the  
effective action has been crucial. In fact, all the terms in ${\cal
L}_\Lambda^F$, except  
for $J(a)$, have the same functional dependence of the brane part of the  
action. Using the freedom to perform finite renormalizations  
we have absorbed in the counterterms everything apart from $J(a)$. 
This leaves the following expression for the renormalized vacuum energy: 
\begin{equation} 
	{\cal L}_{\Lambda}^F = -V_h^R -a^4V_v^R  
	+{(ka)^4 \over 16\pi^2} \int_0^\infty dy y^3 \,
	\ln~\left\{ 1- {K_\mu (y) I_\mu (ay) \over K_\mu
	(ay)  
	I_\mu (y)}\right\}~. 
	\label{rve} 
\end{equation} 
The result is plotted in figure (\ref{fig1}). It qualitatively resembles the
result for massless fermions given by Garriga at al. \cite{Garriga}. Note
that replacing $\mu$ by $-\mu$ leaves the result unchanged appart from a
shift in $V_v^R$, and therefore results with type $II$ boundary conditions
with mass $m$ are equivalent to results for type $I$ boundary conditions
with mass $m-k$.

 
\section{Gauge symmetry and boundary conditions} 
 
In the following we extend the previous considerations in order  
to discuss the most general boundary conditions consistent with the
orbifold symmetry and  
the homogeneity of the spacetime. The interplay between gauge invariance
and boundary conditions is also investigated. 
Once the boundary conditions are specified, the vacuum energy for a
fermion and scalar multiplet,  
described by  
\begin{equation} 
	S=S_{S}+S_{F}~, 
	\label{action} 
\end{equation} 
where  
\begin{equation} 
	S_{S} = {1\over 2} \int d^4x dy |g|^{1\over 2} \left\{ 
	g^{\mu \nu}\partial_\mu \Phi_I^* \partial_\nu \Phi_I  - m_S^2
	\Phi^*_I\Phi_I - \xi R^2 \Phi^*_I \Phi_I
	\right\}~, 
	\label{scalars} 
\end{equation} 
\begin{equation} 
	S_{F} = \int d^4x dy |g|^{1\over 2} \left\{ 
	i \bar{\Psi}_I \underline{\gamma}^M D_M \Psi_I	- m_F ~\epsilon(y)
	\bar{\Psi}_I \Psi_I 
	\right\}~, 
	\label{fermions} 
\end{equation} 
is computed for a variety of boundary conditions. We use the label $I$
to  
index the field multiplets. 
 
\subsection{Homogeneous boundary conditions} 
 
It was pointed out many years ago that the boundary conditions of a
quantum field become very  
rich as soon as the spacetime upon which it is based has a non-trivial
topological structure  
\cite{higuchi,hoso2}. Specifically, if the spacetime is multiply
connected, the fields need not be single valued,  
being, in fact, required to obey weaker boundary conditions.  
It was also noted that the homogeneity of the spacetime and gauge
symmetries produced non trivial constraints  
on the boundary structure of the fields. 
 
A similar situation occurs in the Randall-Sundrum model, where the
boundary conditions can be altered  
from the ones considered in the previous section, according to the
symmetries of the action. 
 
The boundary conditions have to be specified in order to relate the
fields at identified points along the extra dimension and 
since they have to ensure that the action is single valued, we have the
freedom to impose  
weaker boundary conditions on the fields. We assume that the manifold
upon which the quantum field theory is  
homogeneous, meaning that that the physics is the same at every point. 
 
The most general boundary conditions we can write are 
\begin{equation} 
	\Phi_I (x_{\mu} , -y ) = \Sigma_{IJ}^{b} \Phi_J (x_{\mu} , y )~, 
	\label{zs} 
\end{equation} 
\begin{equation} 
	\Psi_I^{\sigma} (x_{\mu} , -y) = \Delta_{IJ}^{b} \L^{\sigma
	\rho}_{b} \Psi_J^{\rho}  
	(x_{\mu} , y )~. 
	\label{zf} 
\end{equation} 
$b = v,~h$, means that we can have different boundary conditions at the
two  
branes\footnote{This is a very important difference with respect to  
\cite{higuchi,hoso2}, in which the manifold $S^1$ did not have  
boundaries}. $\Sigma$ and $\Delta$ are global gauge transformations and
$\Delta^2=1$, $\Sigma^2 =1$. 
The requirement that the action is invariant under (\ref{zs}), (\ref{zf})
results in the following conditions: 
\begin{equation} 
	\Delta_{IJ}^{b*} \Delta_{JK}^b = \delta_{IK}~, 
	\label{a1} 
\end{equation} 
\begin{equation} 
	\L^{\dagger} \gamma^0 \gamma^\mu \L = \gamma^0 \gamma^\mu~, 
	\label{a2} 
\end{equation} 
\begin{equation} 
	\L^{\dagger} \gamma^0 \L = - \gamma^0~, 
	\label{a3} 
\end{equation} 
\begin{equation} 
	\Sigma_{IJ}^{b*} \Sigma_{JK}^b = \delta_{IK}~. 
	\label{a4} 
\end{equation} 
Relations (\ref{a2}), (\ref{a3}) are satisfied uniquely by choosing  
\begin{equation} 
	\L = \pm \gamma^5~. 
	\label{g5} 
\end{equation} 
These are a generalization of the results for $B$ we wrote down earlier. 
The boundary conditions can be further constrained when taking into
account  
the gauge symmetries of the theory. If the theory is gauge invariant,
it would be expected that fields in the same  
gauge equivalence class satisfy the same boundary conditions, which,
in other  
words, means that boundary conditions should be preserved under a gauge  
transformation. To exploit this, insert at each brane a gauge
transformation $U \in G$, where $G$ is the gauge group: 
\begin{equation} 
	\Phi_I^{\prime}(x_\mu, y) =U_S^b(x_\mu,y)\Phi_I(x_\mu, y)~, 
\end{equation} 
\begin{equation} 
	\Psi_I^{\prime}(x_\mu, y) =U_F^b(x_\mu,y)\Psi_I(x_\mu, y)~. 
\end{equation} 
Requiring that the primed fields satisfy the same boundary conditions
as the  
unprimed fields together with the requirement that the gauge
transformation  
be single-valued gives: 
\begin{equation} 
	[U^b_S(x_\mu , y),\Sigma^b]=0~, 
    \label{sb1}  
\end{equation} 
\begin{equation} 
	[U^b_F(x_\mu , y),\Delta^b]=0~. 
    \label{sb2}  
\end{equation} 
Relations (\ref{sb1}), (\ref{sb2}) represent the symmetries of the
boundary conditions  
\cite{hoso2}. 
 
Some comments are now in order. We have seen that the boundary
conditions  
the fields are required  
to obey are weaker the larger is the symmetry of  
the theory and are constrained by the gauge invariance and
${Z}_2-$symmetry  
of the action. Among these constraints  
are the commutation relations given above. These commutation relations
cannot  
be satisfied for any choice of the matrices $\Delta$ and $\Sigma$,
meaning that if we want to keep  
the boundary conditions general, we have to restrict the original
symmetry  
of the theory, which in turn means breaking the gauge symmetry at
classical level.  
If we want to maintain the original symmetry of the classical theory,  
we are forced to restrict the matrices $\Delta$ and $\Sigma$ in order
to satisfy the constraints  
found, i.e. they must belong to the center of the gauge group $G$. 
 
It is instructive to see how this work in simple cases. If we consider a  
single scalar field, relation (\ref{a4}) implies  
that $\Sigma = \pm 1$ (the commutation relations are trivially satisfied  
in this case). This gives: 
\begin{equation} 
	\phi(x_\mu , -y) = \pm \phi(x_\mu , y)~, 
\end{equation} 
where the $+$ sign gives the untwisted field configuration considered
in \cite{Garriga,DJTPLB,GR,flachi,FMTPLB}, 
and the $-$ sign gives the twisted configuration considered in  
\cite{flachi,FMTPLB}. In the single fermion case $\L = \pm \gamma^5$  
and $\Delta = e^{i \theta}$. The boundary conditions are then  
\begin{equation} 
	\Psi(x_\mu , -y) = \pm e^{i \theta} \Psi(x_\mu , y)~, 
\end{equation} 
and the condition $\Delta^2=1$ implies $\theta = 0$, which gives the
fields configurations considered in the  
previous section.  
 
 
\subsection{Effective action} 
 
In this section we will evaluate the effective action for situations in
which a richer boundary structure is possible. 
 
A first simple choice is to take a single real scalar field obeying
different boundary conditions at the two branes.  
Two possibilities arise: the field is even at $y=0$ and odd at $y=\pi r$
or viceversa.  
We call these two cases `twisted' and label them $T_I$ and $T_{II}$
respectively.  The boundary conditions can be applied in a straighforward
manner giving for the eigenvalue equation  
(the functions $j_\nu,~y_\nu,~i_\nu,~k_\nu$ in the scalar case are the
ones  defined in \cite{flachi}): 
\begin{equation} 
	j_\nu(a x_n) Y_\nu(x_n) - y_\nu(a x_n) J_\nu(x_n) = 0~, 
\end{equation} 
for type $T_I$ and  
\begin{equation} 
	j_\nu(x_n) Y_\nu(a x_n) - y_\nu(x_n) J_\nu(a x_n) = 0~, 
\end{equation} 
for type $T_{II}$. The computation of the vacuum energy is no different
from the previous case and the  
renormalized quantum  
corrections can be written as 
\begin{equation} 
	{\cal L}_\Lambda^S = {(ka)^4 \over 64\pi^2} \int dy y^4
	g_\nu^{(T_I,~T_{II})} (y)~, 
	\label{87} 
\end{equation} 
where 
\begin{equation} 
	g_\nu^{T_I} (y) = 1- {i_\nu(ay)K_\nu(y) \over k_\nu(ay)
	I_\nu(y)}~, 
\end{equation} 
and 
\begin{equation} 
	g_\nu^{T_{II}} (y) = 1- {k_\nu(y)I_\nu(ay) \over i_\nu(y)
	K_\nu(ay)}~. 
\end{equation} 
Another simple possibility is to consider a single fermion field obeying
different boundary conditions at the two branes.  
There are two possibilities: 
\begin{equation} 
	\Psi (x_\mu , -y) = +\gamma^5 \Psi (x_\mu , y)~~~~\mbox{at
	$y=0$}~, 
\end{equation} 
\begin{equation} 
	\Psi (x_\mu , -y) = -\gamma^5 \Psi (x_\mu , y)~~~~\mbox{at $y=\pi
	r$}~, 
\end{equation} 
and the reversed one: 
\begin{equation} 
	\Psi (x_\mu , -y) = -\gamma^5 \Psi (x_\mu , y)~~~~\mbox{at
	$y=0$}~, 
\end{equation} 
\begin{equation} 
	\Psi (x_\mu , -y) = +\gamma^5 \Psi (x_\mu , y)~~~~\mbox{at $y=\pi
	r$}~. 
\end{equation} 
The mass eigenvalue equation is 
\begin{equation} 
	J_{-\mu} (ax_n) j_{\mu} (x_n) - J_{\mu} (ax_n) j_{-\mu} (x_n)=0~, 
\end{equation} 
with $\mu = 1/2 +\nu$ for the first case and $\mu = \nu - 1/2$ for the
second case. 
For convenience, we have defined 
\begin{equation} 
	j_{\mu}(z) = \left({1\over 2} - \nu \right)J_\mu(z)+zJ_\mu'(z)~, 
\end{equation} 
\begin{equation} 
	k_{\mu}(z) = \left({1\over 2} - \nu \right)K_\mu(z)+zK_\mu'(z)~. 
\end{equation} 
The evaluation of the vacuum energy goes along the same lines as before  
and the renormalized contribution is found to be: 
\begin{equation} 
	{\cal L}_{\Lambda}^F = {(ka)^4 \over 16\pi^2} \int_0^\infty dy
	y^3\,
	ln~\left\{ 1- {k_\mu (y) I_\mu (ay) \over K_\mu (ay)
	i_\mu (y)}\right\}~. 
	\label{rvetwist} 
\end{equation} 
 
The problem of computing the radiative corrections becomes more
complicated,  
when a gauge symmetry is considered, due to the enlarged  
complexity of the boundary conditions. We have seen that, in order to
maintain the gauge symmetry at classical  
level, the boundary conditions have to satisfy certain symmetries,  
specifically $\Delta$ and $\Sigma$ have to belong to the center of the
gauge group $G$. 
 
As an example, let us consider $G=SU(N)$. $\Delta$ belongs to the center
of $SU(N)$. This modifies the  
boundary conditions in a simple way: the boundary conditions that each
fermion component has to obey can be of type $I$,  
$II$, $T_I$ or $T_{II}$. This can be incorporated in the evaluation of
the vacuum energy in a straightforward manner, giving: 
\begin{equation} 
	{\cal L}_\Lambda^F = \sum_{\kappa \in {(I,~II,~T_I,~T_{II})}}
	N_\kappa^F {\cal L}_\Lambda^F  
	(\kappa)~, 
	\label{fve} 
\end{equation} 
where $N_\kappa^F$ represents the number of components satisfying type
$\kappa$ boundary conditions and ${\cal L}_\Lambda^F (\kappa)$  
is the vacuum energy for each component obeying type $\kappa$ boundary
conditions. 
 
Another simple example is to consider an $SO(N)$ scalar theory, for
which the situation is similar to the previous  
one, giving: 
\begin{equation} 
	{\cal L}_\Lambda^S = \sum_{\kappa \in {(I,~II,~T_I,~T_{II})}}
	N_\kappa^S {\cal L}_\Lambda^S (\kappa)~, 
	\label{sve} 
\end{equation} 
where $N_\kappa^S$ represents the number of scalar components satisfying
type $\kappa$ boundary conditions and ${\cal L}_\Lambda^S (\kappa)$  
is the vacuum energy for each component obeying type $\kappa$ boundary
conditions. 
 
All this can be generalised to a general gauge group whose center is
trivial. In this case the boundary conditions scalar and fermion fields
ought to satisfy 
are still of the same type as before, giving: 
\begin{equation} 
	{\cal L}_\Lambda = {\cal L}_\Lambda^S + {\cal L}_\Lambda^F =
	\sum_{\kappa \in  
	{(I,~II,~T_I,~T_{II})}}  
	\left\{  
	N_\kappa^F {\cal L}_\Lambda^F(\kappa) + N_\kappa^S {\cal
	L}_\Lambda^S (\kappa) 
	\right\}~. 
	\label{tve} 
\end{equation} 
 
 
\section{Conformally coupled case} 
 
The massless, conformally coupled case (studied in \cite{Garriga} for
untwisted field configurations and in  
\cite{FMTPLB} for both the twisted and untwisted case) is worth of some
special  
attention and provides a useful check on the general method used in the  
previous sections.  
 
For type $I$ and type $II$ boundary conditions $\mu = 1/2$, (\ref{53})
can be expressed in terms of elementary functions  
and the integrals are now evaluated at ease, giving for the renormalized  
one-loop vacuum energy: 
\begin{equation} 
	{\cal L}_\Lambda^F = -{3 k^4 a^4 \over 128 \pi^2} \zeta(5)
	(1-a)^{-4}~. 
    \label{v0j} 
\end{equation} 
Similarly, for type $T_I$ and $T_{II}$ boundary conditions $\mu = 1/2$
and  
a straightforward calculation of (\ref{87}) leads to 
\begin{equation} 
	{\cal L}_{\Lambda}^F = {15 \over 16} {3 \over 128 \pi^2}{k^4
	a^4 \over (1-a)^4}\zeta(5)~. 
	\label{rvetwistconf}	 
\end{equation} 
The previous results can be also dealt with by direct summation of the  
mass eigenvalues $m_n$, which are, in general, given by 
\begin{equation} 
	m_n = { k a \over 1-a} (n \pi \pm \theta)~, 
    \label{mth} 
\end{equation} 
where $\theta = 0$ gives the untwisted field configuration, and $\theta
=  
{\pi \over 2}$ gives twisted one. 
The sum over the modes in ${\cal L}_\Lambda^F$ can be performed by using
the properties of the $\zeta -$function  
and without the need of any renormalization: 
\begin{equation} 
     {\cal L}_\Lambda^F=- \lim_{D\rightarrow 4} {1\over 32 \pi^2} {(\pi
     k a)^D\over (1-a)^D}  
     \Gamma(-D/2) \left( \sum_{n=1}^{\infty} (n+\theta )^D +
     \sum_{n=1}^{\infty}  
     (n-\theta)^D\right)~. 
\end{equation} 
The previous result can be expressed in terms of the Hurwitz $\zeta -$
function: 
\begin{equation} 
     {\cal L}_\Lambda^F=- \lim_{D\rightarrow 4} {1\over 32 \pi^2} {(\pi
     k a)^D\over (1-a)^D}  
     \Gamma(-D/2) \left( \zeta_H(-D, \theta) + \zeta_H(-D,  
     1-\theta) \right)~, 
\end{equation} 
which, by using basic properties of $\zeta_H$, can be recast in the form 
\begin{equation} 
     {\cal L}_\Lambda^F=- {3\over 128 \pi^2}{(ka)^4 \over (1-a)^4} \sum  
     {\cos 2n\theta \over n^5}~.  
     \label{th}  
\end{equation} 
Immediate inspection of (\ref{th}) reproduces (\ref{v0j})  
for $\theta =0$ and (\ref{rvetwistconf}) for $\theta =\pi /2$, as it
must  
be. 
 
 
\section{Topological symmetry breaking} 
 
An interesting feature of the boundary conditions on the branes is the 
possibility of breaking bulk gauge symmetries. The residual 
symmetries are those which commute with the two matrices $\Sigma^h$ and 
$\Sigma^v$ introduced in section 3.1. The symmetry breaking mechanism is 
similar to the Wilson-loop symmetry breaking mechanism in non-simply
connected 
spacetimes \cite{higuchi,hoso2,tomstsb,fordtsb,dow1,mign}. 
 
There are two equivalent ways to describe this type of symmetry breaking. 
The non-trivial boundary conditions are useful for evaluating and
comparing 
the effective action for different symmetry breaking schemes, as we
shall do 
below. Alternatively, it is possible to simplify the boundary conditions
by 
performing a gauge transformation which introduces a pure gauge 
field stretching between the two branes. This `Wilson 
line' is the analog of the Wilson loop in the Wilson loop mechanism. If
the 
field strength vanishes, the line integral of the gauge field along a
loop is 
conserved. This need not be true for the Wilson line and the symmetry 
breaking becomes associated with a set of moduli fields. 
 
We shall concentrate on the possible symmetry breaking schemes fields
with 
an $SU(N)$ gauge symmetry. The matrices $\Sigma^h$ and $\Sigma^v$ satisfy 
$(\Sigma^h)^2=(\Sigma^v)^2=I$, the unit matrix. By considering the
eigenspaces 
of $[\Sigma^h,\Sigma^v]^2$, it is easy to show that there is a basis in
which 
the matrices take the block-diagonal form 
\begin{eqnarray} 
\Sigma^h&=&{\rm diag}\{\pm 1,\dots,\mp 1,\sigma_3,\dots,\sigma_3\}~,\\ 
\Sigma^v&=&{\rm diag}\{\pm 1,\dots,\pm 1, 
\sigma_{\theta_1},\dots,\sigma_{\theta_n}\}~,\label{bd} 
\end{eqnarray} 
where $\sigma_1$, $\sigma_2$, $\sigma_3$ are the Pauli matrices and 
\begin{equation} 
\sigma_\theta=\sigma_3\cos 2\theta+\sigma_1\sin 2\theta~. 
\end{equation} 
The residual symmetry group is then 
\begin{equation} 
SU(n_1)\times SU(n_2)\dots\times SU(n_p)\times U(1)^q. 
\end{equation} 
There are $SU(n)$ factors for each of the four combinations of the $\pm1$ 
entries along the diagonals of the matrices and further $SU(n)$ factors
for 
each repeated value of $\theta$. 
 
For example, if $G$ is the group $SU(5)$, we can take 
\begin{eqnarray} 
\Sigma^h&=&{\rm diag}\{-1,-1,-1,\sigma_3\}~,\\ 
\Sigma^v&=&{\rm diag}\{-1,-1,-1,\sigma_{\theta}\}~. 
\end{eqnarray} 
The group reduces to $G\to SU(4)\times U(1)$ if $\theta=0$ or $\pi/2$
and $G\to 
SU(3)\times U(1)$ otherwise. 
 
The action (\ref{scalars}) or (\ref{fermions}) splits into separate terms,
with one 
term for each of the block diagonal entries (\ref{bd}). Each term gives a 
contribution to the effective potential. The $\pm 1$ entries correspond
to the type $I$ and $II$, twisted and untwisted boundary conditions considered
in section 3.2. The $\sigma_\theta$ entries correspond to the following
boundary 
conditions on the fermion modes at the hidden brane $y=0$ and the visible 
brane $y=r\pi$, 
 \begin{eqnarray} 
G_R(0)&=&\sigma_3 G_R(0)~,\\ 
G_R'(0)&=&-\sigma_3 G_R'(0)~,\\ 
G_R(r\pi)&=&\sigma_\theta G_R(r\pi)~,\\ 
G_R'(r\pi)&=&-\sigma_\theta G_R'(r\pi)~. 
\end{eqnarray} 
The boundary conditions for $G_L$ modes and scalar field modes have an
equivalent form. 
 
The fermion mode functions were given in equations (\ref{gggl}) and 
(\ref{gggr}). Substituting these modes into the boundary conditions
gives the 
values for the masses $m_n$. Introduce 
\begin{eqnarray} 
p_\nu&=& J_\nu\left({m_n\over k}\right)Y_\nu\left({m_n\over ka}\right) 
-J_\nu\left({m_n\over ka}\right)Y_\nu\left({m_n\over k}\right)~,\\ 
q_\nu&=& J_\nu\left({m_n\over k}\right)Y'_\nu\left({m_n\over ka}\right) 
-J_\nu\left({m_n\over ka}\right)Y'_\nu\left({m_n\over k}\right)~,\\ 
r_\nu&=& J'_\nu\left({m_n\over k}\right)Y_\nu\left({m_n\over ka}\right) 
-J'_\nu\left({m_n\over ka}\right)Y_\nu\left({m_n\over k}\right)~,\\ 
s_\nu&=& J'_\nu\left({m_n\over k}\right)Y'_\nu\left({m_n\over ka}\right) 
-J'_\nu\left({m_n\over ka}\right)Y'_\nu\left({m_n\over k}\right). 
\end{eqnarray} 
The values of $m_n$ are given by 
\begin{equation} 
p_{\nu+1/2}s_{\nu+1/2}\cos^2\theta+ 
q_{\nu+1/2}r_{\nu+1/2}\sin^2\theta=0. 
\label{108} 
\end{equation} 
For $\theta=0$, this reduces to the previous cases $p_{\nu+1/2}=0$
(for type 
$I$ boundary conditions) and $s_{\nu+1/2}=0$ (for type $II$
boundary conditions). 
 
In the massless fermion and the conformally invariant scalar field
theories 
the Bessel functions become trigonometric functions and the values of
$m_n$ 
are given by (\ref{mth}). The vacuum energy from one $\theta$ `block' 
can be expressed in terms of Hurwitz zeta functions and evaluates to 
\begin{equation} 
{\cal L}_{V}=\pm {3gk^4 a^4\over 128\pi^2}(1-a)^{-4} 
\sum_{n=1}^\infty{\cos(2n\theta)\over n^5} 
\end{equation} 
where the upper sign is for fermions, the lower for bosons and $g$ is the 
dimension of the fermion representation. The vacuum energy is extremised 
for $\theta=0$ and $\theta=\pi/2$, where the result reduces to the 
untwisted and twisted results respectively. This remains the case for
massive fields because of the symmetries $\theta \rightarrow -\theta$ and
$\theta\rightarrow \pi -\theta$ in the formula for the $m_n$ (\ref{108}).

The full effective action will include, not only potential terms, but 
also induced kinetic terms for the moduli fields. Their presence can be 
inferred from the form of the $a_{5/2}$ heat kernel coefficient, which 
depends on the boundary conditions and includes derivatives of the 
matrices $\Sigma^h$ and $\Sigma^v$. The symmetry breaking mechanism is 
therefore truly dynamical, and differs from the usual Wilson loop 
mechanism in this respect. The closest analogy is to the symmetry breaking
associated with quantum wormholes \cite{mign}.
 
 
\section{Radius stabilization and self-consistent solutions} 
 
In a previous paper \cite{flachi}, by looking at the quantum corrected
Einstein equations,  
we have examined the conditions for self-consistency of the
Randall-Sundrum spacetime  
and obtained a self-consistency relation, which the radius had to
satisfy.  
Specifically, if 
\begin{equation} 
	\Gamma = S_G - \int dv_x~F(a)~, 
\end{equation} 
denotes the full quantum corrected effective action, the requirements are 
\begin{equation} 
	\left({\delta \Gamma \over \delta g_{\mu \nu}} \right)_{g_{\mu
	\nu}=\eta_{\mu \nu}} = 0~, 
\end{equation} 
\begin{equation} 
	\left({\delta \Gamma \over \delta a}\right)_{g_{\mu \nu}=\eta_{\mu
	\nu}} = 0~. 
\end{equation} 
The first forces the Randall-Sundrum  
solution to be a solution of the quantum corrected Einstein's equations,
the second being a requirement for the radius to be an extremum of the  
effective potential. When quantum corrections are included 
\begin{equation} 
	F(a) = V_h +a^D V_v + f(a)~, 
\end{equation} 
the following relation is obtained 
\begin{equation} 
	0= D a^D (V_v +V_h) + (1-a^D)af'(a) + Daf(a)~. 
\end{equation} 
In \cite{flachi}, we studied these consistency requirements for quantized
bulk  
scalar fields. In that case, the conclusion was that when the balancing  
condition between the brane tension is forced to hold at quantum level
also,  
quantum effects were offering no self-consistent radius stabilization  
mechanism. When the balancing condition was relaxed, it was possible to  
find self-consistent solutions only at the price of fine tuning the
brane  
tensions to a considerable degree. This was found to be in agreement
with  
\cite{Garriga,FMTPLB}. 
 
Including fermions in the analysis might give some hope to find a  
self-consistent solution with a less degree of fine tuning. 
If one defines  
\begin{equation} 
	f(a) =	{\cal L}_\Lambda^S  + {\cal L}_\Lambda^F~, 
    \label{cr}  
\end{equation} 
one finds that, if $V_v+V_h = 0$ is required to hold, than there is no
self-consistent solution for which  
the hierarchy problem is simultaneously solved. If this  
condition is relaxed then a severe fine tuning of the brane tensions is
still  
required in order to satisfy (\ref{cr}). Unfortunately, one has to resort
to a numerical approach to verify this,  
nevertheless in some special cases it is possible to see this
analytically.  
For example in the massless case one has ($c_s$ and $c_f$ are irrelevant  
numerical factors): 
\begin{equation} 
	0 =  (V_v+V_h)(1-a)^4 + (c_s-c_f)(1+a^2-a^6)~. 
    \label{cr2}  
\end{equation} 
It is now straighforward to see that in order to have a solution to  
(\ref{cr2}) and simultaneously solve the hierarchy problem ($a \simeq  
10^{-18}$) a dramatic fine tuning of the parameters in (\ref{cr2})
is required. 
 
 
\section{Conclusions} 
 
We have evaluated the one-loop radiative corrections to the effective
action arising from a massive bulk fermion quantized on the  
Randall-Sundrum background spacetime. As in the scalar field case,
it is not  
possible to obtain an exact result for general curved membranes, but it
is possible to resort to a heat kernel expansion and compute the  
effective action to any desirable order. The vacuum energy density,
which is the first term in the expansion,  
has been recognized to play a role in the Randall-Sundrum model for its
contribution to the stability of the branes separation which  
is related to the gauge hierarchy problem.  
We showed that the fields can obey different boundary conditions from
the ones considered in  
\cite{Garriga,DJTPLB,GR,flachi,FMTPLB} and this gives rise to
modifications in the vacuum energy. 
 
We have clarified the relation between gauge invariance and boundary
conditions and showed how  
these are constrained by the gauge invariance, having to obey to what  
Hosotani called the symmetries of the boundary conditions \cite{hoso2}.  
This richer boundary structure has to be taken into account when scalar
or fermion multiplets are considered and in this case  
the vacuum energy is calculated for $SU(N)$ fermions, $SO(N)$ scalars
and  
generally for situations in which the center of the gauge group is
trivial. 
 
The massless (conformally coupled) case is dealt with by direct summation
of the eigenvalues and as a limiting case of the general result.  
This has provided a check on both our general method and a comparison
with  
previous results. 
 
The possibility of breaking the bulk gauge symmetries by using a mechanism
similar to the Wilson  
loop symmetry breaking in non-simply connected spacetimes has been
analyzed. We concentrated on the  
possible breaking schemes when the gauge group is $SU(N)$ and showed
that the residual symmetry group  
is always of the form $SU(n_1) \times ... \times SU(n_p) \times U(1)^q$. The
vacuum energy depends on a set of moduli fields $\theta$ and is generally
extremised for $\theta=0$ or $\theta=\pi/2$. It would be interesting to
investigate the dynamical implications of these moduli fields. There may also
be a connection with the wormhole symmetry breaking mechanism \cite{mign}. 

The self-consistency, discussed for 
quantized scalar fields \cite{flachi}, has been examined when massive fermion
are included.  It is shown that in this case also it is not possible to
stabilize the radius and simultaneously solving the hierarchy problem without  
a considerable degree of fine tuning, supporting the previous claims of
\cite{Garriga,DJTPLB,GR,flachi,FMTPLB}. 
 
 
\section*{Acknowledgements} 
 
A. Flachi is grateful to the University of Newcastle upon Tyne for the
award of a Ridley Studentship. 
 

\begin{thebibliography}{99} 
 
\bibitem{Kaluza} 
T. Kaluza, Sitzungsber. Preuss. Akad. Wiss. Berlin, Math. Phys. Kl. (1921)
966. 
 
\bibitem{Klein} 
O.~Klein, Z. Phys. {\bf 37} (1926) 895; Nature (London) {\bf 118} 516. 
 
\bibitem{DeWitt} 
B.~S.~DeWitt, in {\bf Relativity, Groups, and Topology}, edited by  
C.~DeWitt and B.~S.~DeWitt (Gordon and Breach, New York, 1964). 
 
\bibitem{Witten} 
E.~Witten, Nucl. Phys. B {\bf 186} (1981) 412. 
 
\bibitem{Nahm} 
W.~Nahm, Nucl. Phys. B {\bf 135} (1978) 149. 
 
\bibitem{DuffNilssonPope} 
M.~J.~Duff, B.~Nilsson, and C.~N.~Pope, Phys. Rep. {\bf 130} (1986) 1 . 
 
\bibitem{Arkani} 
N.~Arkani-Hamed, S.~Dimopoulos, G.~Dvali, Phys. Lett. {\bf B429} (1998)  
263; I.~Antoniadis, N.~Arkani-Hamed, S.~Dimopoulos, G.~Dvali,
Phys. Lett.  
{\bf B436} (1998) 257. 
 
\bibitem{RS} 
L.~Randall and R.~Sundrum, Phys.~Rev.~Lett. {\bf 83} (1999) 3370; ibid.
4690. 
 
\bibitem{GW1} 
W.~D.~Goldberger and M.~B.~Wise, Phys. Rev. {\bf D60} (1999) 107505. 
 
\bibitem{DHR} 
H.~Davoudiasl, J.~L.~Hewett, and T.~G.~Rizzo, Phys. Lett. {\bf B473}
(2000)	43. 
 
\bibitem{Pomarol} 
A.~Pomarol, Phys. Lett. {\bf B486} (2000) 153. 
 
\bibitem{Grossman} 
Y.~Grossman and M.~Neubert, Phys. Lett. {\bf B474} (2000) 361. 
 
\bibitem{Kitano} 
R.~Kitano, Phys. Lett. {\bf B481} (2000) 39. 
 
\bibitem{Chang} 
C.~V.~Chang and J.~N.~Ng, Phys. Lett. {\bf B488} (2000) 390. 
 
\bibitem{Gherghetta} 
T.~Gherghetta and A.~Pomarol, Nucl. Phys. B {\bf 586} (2000) 141. 
 
\bibitem{Changetal} 
S.~Chang, J.~Hisano, H.~Nakano, N.~Okada, and M.~Yamaguchi, Phys. Rev. D  
{\bf 63} (2000) 084025. 
 
\bibitem{DHR2} 
H.~Davoudiasl, J.~L.~Hewett, and T.~G.~Rizzo, Phys. Rev. {\bf D63}
(2001)  
075004. 
 
\bibitem{Huber} 
S.~J.~Huber and Q.~Shafi, Phys. Rev. {\bf D63} (2001) 045010. 
 
\bibitem{FTplb} 
A.~Flachi and D.~J.~Toms, Phys. Lett. {\bf  B491} (2000) 157. 
 
\bibitem{georgi1} 
H. Georgi, A. Grant, G. Hailu, Phys. Rev. {\bf D63} (2001) 2853. 
 
\bibitem{georgi2} 
H. Georgi, A. Grant, G. Hailu, Phys. Lett. {\bf B506} (2001) 207. 
 
\bibitem{lukas} 
A. Lukas, P. Ramond, A. Romanino, G. G. Ross, JHEP {\bf 04} (2001) 010.

\bibitem{FMTPLB} 
A. Flachi, I. G. Moss, D. J. Toms, $hep-th/0103138$.
 
\bibitem{daemi} 
S. Randjbar-Daemi, M. Shaposhnikov, Phys. Lett. {\bf B492} (2000) 361. 
 
\bibitem{gibbons} 
G. Gibbons, R. Kallosh, A. Linde, JHEP {\bf 01} (2001) 022. 
	 
\bibitem{CW} 
P.~Candelas and S.~Weinberg, Nucl. Phys. {\bf B237} (1984) 397. 
 
\bibitem{DJTgravity} 
D.~J.~Toms, Phys. Lett. {\bf B129} (1983) 31. 
 
\bibitem{DJTChalkRiver} 
D.~J.~Toms, in {\bf An Introduction to Kaluza-Klein Theories}, edited by  
H. C. Lee (World Scientific, Singapore, 1984). 
 
\bibitem{ACFbook} 
{\bf Modern Kaluza-Klein Theories}, edited by T. Appelquist, A. Chodos,
P.G.O. Freund  
(Addison-Wesley, 1987). 
 
\bibitem{Garriga} 
J. Garriga, O. Pujol\`{a}s, and T. Tanaka, $hep-th/0004109$. 
 
\bibitem{DJTPLB}  
D.~J.~Toms, Phys. Lett. {\bf B484} (2000) 149. 
 
\bibitem{GR} 
W. D. Goldberger and I. Z. Rothstein, Phys. Lett. {\bf B491} (2000) 339. 
 
\bibitem{flachi} 
A. Flachi and D. J. Toms, $hep-th/0103077$.   
 
\bibitem{ponton} 
E. Ponton, E. Poppitz, $hep-ph/0105021$. 
 
\bibitem{horava} 
M. Fabinger, P. Horava, Nucl. Phys. B {\bf 580} (2000) 243. 
 
\bibitem{Gilkey1} 
P. B. Gilkey, J. Diff. Geom. {\bf 10} (1975) 601. 
 
\bibitem{Gilkey2} 
T. P. Branson, P. B. Gilkey, K. Kirsten, D. V. Vassilevich,
Nucl. Phys. {\bf B563} (1999) 603. 
 
\bibitem{kirsten} 
K. Kirsten, Class. Quantum Grav. {\bf 15} (1998) L5. 
 
\bibitem{Gilkey3} 
P. B. Gilkey, K. Kirsten, D. V. Vassilevich, Nucl. Phys. {\bf B601}
(2001) 125. 
 
\bibitem{mossplb} 
I. G. Moss, Phys. Lett. {\bf B491} (2000) 203. 
 
\bibitem{bordag} 
M. Bordag, D. V. Vassilevich, J. Phys. A {\bf 32} (1999) 8247. 
 
\bibitem{kunst} 
S.R. Huggins, G. Kunstatter, H.P. Leivo, D.J. Toms, Phys. Rev. Lett. {\bf
58}  
(1987) 296; Nucl. Phys. {\bf B301} (1988) 627. 
 
\bibitem{kim1} 
J. E. Kim, B. Kyae and H. M. Lee, Phys. Rev. {\bf D62} (2000) 045013. 
	 
\bibitem{kim2} 
J. E. Kim, B. Kyae and H. M. Lee, Nucl. Phys. {\bf B582} (2000) 296. 
	 
\bibitem{corradini} 
O. Corradini and Z. Kakushadze, Phys. Lett. {\bf B494} (2000) 302. 
	 
\bibitem{nojiri} 
S. Nojiri and S. D. Odintsov, JHEP {\bf 07} (2000) 049. 
	 
\bibitem{navromatos} 
N. Navromatos and J. Rizos, Phys. Rev. {\bf D62} (2000) 124004. 
	 
\bibitem{neupane} 
I. P. Neupane, JHEP {\bf 09} (2000) 040. 
 
\bibitem{low} 
I. Low and A. Zee, Nucl. Phys. {\bf B585} (2000) 395. 
 
\bibitem{parker} 
L. Parker, D. J. Toms, Phys. Rev. {\bf D31} (1985) 953. 
 
\bibitem{kirst1} 
M. Bordag, E. Elizalde, K. Kirsten, J. Math. Phys. {\bf 37} (1996) 895. 
 
\bibitem{kirst2} 
M. Bordag, K. Kirsten, J. S. Dowker, Comm. Math. Phys. {\bf 182} (1996)
371. 
 
\bibitem{higuchi} 
A. Higuchi, L. Parker, Phys. Rev. {\bf D37} (1988) 2853. 
 
\bibitem{hoso2} 
Y. Hosotani, Ann. Phys. {\bf 190} (1989) 233; Phys. Lett. {\bf B126}
(1983) 309. 
 
\bibitem{tomstsb} 
D. J. Toms, Phys. Lett. {\bf B126} (1983) 445. 
 
\bibitem{fordtsb} 
L. H. Ford, Phys. Rev. {\bf D21} (1980) 933. 
 
\bibitem{dow1} 
J. S. Dowker, S. Jadhav, Phys. Rev. {\bf D39} (1989) 1196; ibid. 1196. 
 
\bibitem{mign} 
S. Mignemi, I. G. Moss, Phys. Rev. {\bf D48} (1993) 3725. 
 
\end{thebibliography}
\end{document}